\documentclass[lettersize,journal]{IEEEtran}
\usepackage{amsmath,amsfonts}
\usepackage{algorithmic}
\usepackage{algorithm}
\usepackage{array}
\usepackage[caption=false,font=normalsize,labelfont=sf,textfont=sf]{subfig}
\usepackage{textcomp}
\usepackage{tabularx}
\usepackage{stfloats}
\usepackage{url}
\usepackage{verbatim}
\usepackage{graphicx}
\usepackage{cite}
\usepackage[linewidth=1pt]{mdframed}
\newtheorem{theorem}{Theorem}
\usepackage{amssymb}
\newcommand\qedsymbol{$\blacksquare$\vspace{1mm}}

\begin{document}

% \title{Securing IoT Authentication against Modeling Attacks by PUF-Protocol Co-Design}

\title{A lightweight PUF-based authentication protocol}
\author{Yu Zhuang, \IEEEmembership{IEEE Member,} and Gaoxiang Li%, \IEEEmembership{IEEE Student Member}
\thanks{Manuscript received Month XX, 2024; revised Month XX, 2024.}
\thanks{This work was supported in part by the National Science Foundation under grant No. 2103563 (Corresponding author: Yu Zhuang).}
\thanks{Yu Zhuang is with Department of Computer Science, Texas Tech University, Lubbock, TX 79409 USA (e-mail: yu.zhuang@ttu.edu).}
\thanks{Gaoxiang Li is with Department of Computer Science, Texas Tech University, Lubbock, TX 79409 USA (e-mail: gaoli@ttu.edu).}
}

% The paper headers
\markboth{IEEE Transactions on Computers,~Vol.~XX, No.~XX, Month~20XX}% 
{Shell \MakeLowercase{\textit{Zhuang et al.}}: Securing IoT Authentication against Modeling Attacks by PUF-Protocol Co-Desig}

\IEEEpubid{0000--0000/00\$00.00~\copyright~2024 IEEE}
% Remember, if you use this you must call \IEEEpubidadjcol in the second
% column for its text to clear the IEEEpubid mark.

\maketitle

\begin{abstract}
Lightweight authentication is essential for resource-constrained Internet-of-Things (IoT). Implementable with low resource and operable with low power, Physical Unclonable Functions (PUFs) have the potential as hardware primitives for implementing lightweight authentication protocols. The arbiter PUF (APUF) is probably the most lightweight strong PUF capable of generating exponentially many challenge-response pairs (CRPs), a desirable property for authentication protocols, but APUF is severely weak against modeling attacks. Efforts on PUF design have led to many PUFs of higher resistance to modeling attacks and also higher area overhead. There are also substantial efforts on protocol development, some leverage PUFs' strength in fighting modeling attacks, and some others employ carefully designed protocol techniques to obfuscate either the challenges or the responses with modest increase of area overhead for some or increased operations for some others. To attain both low resource footprint and high modeling attack resistance, in this paper we propose a co-design of PUF and protocol, where the PUF consists of an APUF and a zero-transistor interface that obfuscates the true challenge bits fed to the PUF. The obfuscated PUF possesses rigorously proven potential and experimentally supported performance against modeling attacks when a condition is met, and the protocol provides the condition required by the PUF and leverages the PUF's  modeling resistance to arrive at low resource overhead and high operational simplicity, enabling lightweight authentications while resisting modeling attacks.

% Lightweight authentication is crucial for resource-constrained Internet-of-Things (IoT) devices. Physical Unclonable Functions (PUFs), particularly capable of low resource and low power operation, offer a promising solution as hardware primitives for such protocols. The arbiter PUF (APUF) is probably the most lightweight strong PUF capable of generating exponentially many challenge-response pairs (CRPs), a desirable property for authentication protocols, but APUF is severely weak against modeling attacks. Despite advancements in PUF designs that resist these attacks with increased complexity and area overhead, and protocol improvements that obfuscate challenges or responses, challenges remain in balancing attack resistance with resource efficiency. To attain both low resource footprint and high modeling attack resistance, in this paper we propose a co-design of PUF and protocol, where the PUF consists of an APUF and a zero-transistor interface that obfuscates the true challenge bits fed to the PUF. The obfuscated PUF possesses rigorously proven potential and experimentally supported performance against modeling attacks when a condition is met, and the protocol provides the condition required by the PUF and leverages the PUF's modeling resistance to arrive at low resource overhead and high operational simplicity, enabling lightweight authentications while resisting modeling attacks. 
 
\end{abstract}


\section{Introduction}\label{sec1}

\IEEEPARstart{P}{hysical} Unclonable Functions (PUFs) are hardware primitives for implementing security protocols. Small scale variations of integrated circuits exist in fabricated silicon chips. These variations are regarded as side effects for conventional circuits \cite{ruhrmair2014pufs, suh2007physical}, but they make each chip unique and can be exploited to prevent semiconductor re-fabrication. PUFs utilize these variations to produce responses unique for individual  circuits  \cite{gassend2002controlled,gassend2002silicon, suh2007physical, herder2014physical, ruhrmair2014pufs, Yu2016lockdown}, and hence are not physically reproducible. These physical variations are hardware fingerprints that can be used for security purposes. Instead of storing secret keys in nonvolatile memories, PUFs retrieve the secret information of physical variations to produce unique responses as signatures of silicon chips. Implementable with simplistic circuits with thousands of transistors, PUFs incur low area overhead and energy consumption, rendering them potential candidates for resource-constrained IoT devices.

Though physically unclonable, PUFs face non-physical threats from modeling attacks. Many PUF-based protocols use open communication channels for sending-receiving challenge response pairs (CRPs) of the PUF, since such channels incur lower implementation cost than secure  communication channels. The CRPs used in authentications can be collected by attackers to build machine learning models that enable attackers to develop software to impersonate the PUF.

Delay-based PUFs, leveraging different gate delays at multiple locations, can create exponentially many combinations of delay differences to produce circuit-dependent responses, leading to exponentially many CRPs. PUFs that admit exponentially many CRPs are called strong PUFs, which are essential for implementing authentication protocols that need to execute more than a small number of authentications in their operation lifespan. In addition, the huge number of CRPs can generate an unlimited supply of ``keys", enabling each to be used only once without having to be stored in devices' memories. Due to the unlimited CRPs, PUF-based protocols usually do not encrypt CRPs, in order to reduce resource overhead for implementing encryption on IoT devices. The un-encrypted CRPs sent over open communication channels can be collected by attackers who use the CRPs to train machine learning models to predict responses to challenges, giving rise to what is called modeling attacks on strong PUFs.
\IEEEpubidadjcol

The Arbiter PUF (APUF) \cite{gassend2004identification,lee2004technique} is probably is most lightweight delay-based strong PUFs with an area overhead of about 6\,n gate-equivalents (GEs) for an $n$-bit APUF, but is severely vulnerable to modeling attacks.  Efforts to improve modeling-attack-resistance over the APUF have resulted in many sophisticated PUF designs, e.g. the XOR PUF (XPUF) \cite{suh2007physical}, the FFPUF \cite{gassend2004identification,lee2004technique,lim2005extracting}, the LSPUF \cite{majzoobi2008lightweight}, the IPUF \cite{nguyen2019interpose}, the LPPUF\cite{Wisiol2022Towards}. These PUFs improved security against modeling attacks, but all have area overhead multiple times that of the APUF.  Many of them still succumb to modeling attacks \cite{ruhrmair2010modeling, Santikellur2019deep, Wisiol2022IEEE, Wisiol2022Towards} unless large circuit architecture sizes are adopted, e.g. 7 or more component APUFs for a 64-bit LSPUF \cite{Wisiol2019Breaking, Santikellur2019deep}, over 10 component PUFs for a 64-bit XPUF \cite{Wisiol2022IEEE}, and over 9 component APUFs for a 64-bit IPUF \cite{Thapaliya2021Machine}. Large circuit architecture sizes obviously lead to higher area overhead. For instance, a PUF with 8 component APUFs has an area overhead 8 times that of the APUF. In addition, some of PUFs also succumb to reliability machine learning attacks \cite{Tobisch2021Combining} even with large circuit architecture sizes.

Besides sophisticated PUFs, there have also been protocol-level efforts to fight modeling attacks. One approach is to use highly attack-resistant PUFs as is done by the Lockdown Protocol II in \cite{Yu2016lockdown}. Another approach is to obfuscate the challenge  \cite{ye2016rpuf,Gao2016Obfuscated,Zalivaka2019Reliable}, and a more broadly used approach is to obfuscate or hide the responses   \cite{Majzoobi2012Slender, Yu2014Noise, Gu2021Modeling, Chen2022Novel}. We also saw two protocols \cite{Zhang2021Set_based,Idriss2021Lightweight} that obfuscates both the challenge and the responses.

To defend against modeling attacks while maintaining low hardware overhead, in this paper we introduce a highly simplistic mutual authentication protocol which is itself lightweight in addition to use a highly lightweight PUF. The protocol is tailor-designed to (i) shield a weak spot of the PUF from being attacked and (ii) utilize the strength of the PUF in resisting modeling attacks.

The protocol employs an APUF equipped with a zero-transistor challenge interface that takes ghost input bits,  bits that are not fed to any PUF stage, leading to more input bits than PUF stages. The ghost bits are inserted into the input vector at random positions with different ghost bit positions for different interface instances, obfuscating the bits fed to the PUF. The challenge interface was proven to turn an APUF into a binary classification problem whose separation surface is defined by a high-order multivariate polynomial with the polynomial order  approximately proportional to the number of ghost bits when a condition on the ghost bits is met. This theoretical result shows that an APUF equipped with ghost bits becomes a highly nonlinear classification problem.  Increased nonlinearity in general leads to lower machine-learnability, and this theoretical analysis has been supported by experimental attacks on instances of APUFs equipped with the challenge interface.

%%%%%%%%%%%%%%%%%%%%%%%%%%%%%%%%%%%%%%%%%%%%%%%%%%%%%%%%%%%%%
\section{The Challenge Obfuscation Interface}\label{sec2}
\subsection{The Design Philosophy}\label{sec2.A}

Feature generation is important for machine learning, and it is especially true for machine learning attacks, or modeling attacks, of PUFs. To the best of our knowledge, all modeling attacks \cite{ruhrmair2010modeling,aseeri2018machine,Wisiol2022IEEE} of delay-based strong PUFs transform the challenge ($c_1,c_2,\cdots,c_n)$ into the feature vector ($\phi_1,\phi_2,\cdots,\phi_n$) according to 
\begin{equation} \label{eq1}
\phi_i = (2c_i-1)(2c_{i+1}-1)\cdots(2c_n-1),
\end{equation} 
before applying machine learning procedures, where $c_i$ is the bit fed to the $i$-th PUF stage. The transform (\ref{eq1}) turns the relation between the response of the APUF and the feature vector ($\phi_1, \phi_2, \cdots, \phi_n$) into a linear classification problem, and turns other delay-based PUFs into more easily learnable classification problems. The work of Xu, et al. \cite{Xu20023Modeling} shows that without the transform (\ref{eq1}), APUF has withstood modeling attacks with one million CRPs. We also tried machine learning methods to attack APUFs with even more CRPs without using the transform (\ref{eq1}), and all our attacks failed, showing the necessity of the feature-generating transform for machine learning attacks of APUFs.

We observed that the feature-generating transform (\ref{eq1}) requires the knowledge of the positions of all challenge bits that are fed to the PUF. Specifically, the generation of feature bit $\phi_i$ needs challenge bits $c_i, c_{i+1},\cdots,c_n$.  Thus, we are motivated to think that if attackers do not know which input bit is fed to which PUF stage, then the feature-generation transform (\ref{eq1}) will not have adequate information to generate the correct feature vector ($\phi_1,\phi_2,\cdots,\phi_n$), and thereby machine learning attacks can probably be defeated. Our approach to obfuscating the positions of challenge bits is to insert ghost bits randomly into the input challenge.

%%%%%%%%%%%%%%%%%%%%%%%%%%%%%%%%%%%%%%%%%%%%%%%%%%%%%%%%%%
\subsection{The Challenge Input Interface}\label{sec2.B}

We propose an interface for the input challenge which, as elaborated below, accepts more input bits than the number of stages of the PUF.

%%%%%%%%%%%%%%%%%%%%%%%%%%%%%%%%%%%%%
\vspace{2mm}
\noindent{\bf The $m$-plus-bits Interface}
\begin{itemize}
\item For a PUF with $n$-stages, the challenge input interface has $(n+m)$ input bits and $n$ output bits. Input bits are denoted by $(b_1,b_2, \cdots, b_{n+m})$ and the output bits $(c_1,c_2, \cdots, c_n)$ with $c_i$ being the bit fed to the $i$-th stage of the PUF.

\item For each PUF instance, $n$ out of the $(n+m)$ input bits are randomly selected, and the remaining $m$ bits, called ghost bits, are not fed to the PUF. Denote the set of ghost bits by $G_m=\{i_1, i_2, \cdots, i_m\}$ with $i_1 < i_2 < \cdots < i_m$.

\item The selected $n$ bits are fed to the PUF circuit in the original order, that is,
\begin{equation}
\label{interface.3}
\left\{
\begin{aligned}
c_i ~ =b_{i} ~ &\mbox{ for }i < i_1, \\
c_{i-j}\!=b_{i}&\mbox{ for }i_j\!<\!i\!<\!i_{j+1},~j\!=\!1,2,\cdots,m\!-\!1, \\
c_{i-m}\!=b_{i} &\mbox{ for }i_m < i .
\end{aligned}
\right .
\end{equation}
\end{itemize}
\vspace{1mm}
%%%%%%%%%%%%%%%%%%%%%%%%%%%%%%%%%%%%%%

The effectiveness of this interface in obfuscating the true challenge bits requires the preclusion of open access to the interfaced PUF, where open access to the PUF means that the PUF returns a response when any (n+m)-bit challenge is fed to it. If the PUF is openly accessible, an attacker can find out the ghost bits by a chosen-challenge attack described below.

\vspace{1mm}
\begin{mdframed}
{\bf\em The Chosen-Challenge Attack}\vspace{1mm}
    
{\em An attacker chooses a small set $S_0$ of random challenges, say 10 challenges. To see if the $i$-th challenge bit is a ghost bit, the attacker constructs a set $S_i$ of challenges which are generated from $S_0$ by flipping the $i$-th bit of all challenges in $S_0$. A challenge from $S_0$ and a challenge from $S_i$ form a corresponding pair if the two challenges differ only in the $i$-th bit. Then, the attacker feeds both sets of challenges to the PUF, and if for all or almost all corresponding pairs of challenges the interfaced PUF produces the same responses, then the $i$-th bit is a ghost bit.}
\end{mdframed}
\vspace{1mm}

Thus, to prevent chosen-challenge attack, open accesses to the PUF must be disabled. Then, an attacker can only passively collect CRPs from eavesdropping on the communications between the PUF and the PUF's legitimate partner server. 
%One way to do this in a protocol is to require at least some of the challenge bits supplied by the PUF-containing device, as done by Protocol II in [lockdown]. 

Then, a natural question is if it is feasible for an attacker to launch chosen-challenge attacks when all CRPs are passively collected. More specifically, if an attacker has millions of CRPs accumulated passively, what is the likelihood that the millions of CRPs contain two small subsets which have the same property as $S_0$ and $S_1$ described in the chosen-challenge attack? A careful examination of the attack led us to the following observation.

\vspace{1mm}
\begin{mdframed}
{\em If all challenges fed to the PUF interface are random in the sense that every bit in every challenge has an equal probability to be 0 or 1, then the probability for an attacker to find two sets $S_0$ and $S_1$ as described in the Chosen-Challenge Attack 
%a corresponding pair of challenges which differ in only one bit
%two attack-enabling sets of challenges 
from 1 million passively accumulated CRPs is lower than $2^{-450}$ for $n=64$ and $m=20$}.
\end{mdframed}
\vspace{1mm}

Let us give the analysis that led to the estimate of probability given above. %to find a pair of  sets $S_0$ and $S_1$ from $N$ passively accumulated CRPs. 
Since each bit is equally likely to be 0 or 1, the probability of having a corresponding pair of challenges which differ in only one bit that are chosen from the $N$ CRPs is $\frac{N(N-1)}{2} 2^{-(n+m)}$, since there are $\frac{N(N-1)}{2}$ possible pairs and the second challenge in the pair has all (n+m) bits fully determined. For $N\!=\!1$ million, $n\!=\!64$ and $m=20$, $\frac{N(N-1)}{2}  2^{-(n+m)} < 2^{-45}$. The probability to have 10 corresponding pairs of challenges in sets $S_0$ and $S_1$ is $\frac{N!}{20!(N-20)!} 2^{-10(n+m)}$, which 
less than $2^{-450}$ for $n=64$, $m=20$, and $N=1$ million. Even if $N=1$ billion, the probability of having a corresponding pair of challenges is less than $2^{-25}$, and the probability to have 10 corresponding pairs of challenges in sets $S_0$ and $S_1$ is less than $2^{-250}$. It can be reasonably assumed that the number of challenges used by a protocol during the entire operating lifespan of the protocol is bounded by 1 billion. Thus, the probability to find a pair of subsets like $S_0$ and $S_1$ from passively accumulated CRPs is close to zero. A chosen-challenge attack needs multiple pairs of sets like $S_0$ and $S_i$ with multiple challenges in each of $S_0$ and $S_i$, and the probability to find multiple pairs of sets like $S_0$ and $S_i$ will be even lower.

While the chosen-challenge attack can completely destroy the challenge obfuscation interface, our observation shows that such attacks can be defeated if all challenge bits fed to the PUF interface are random with equal likelihood of being 0 or 1.%and LFSRs are up to the task for for generating such challenge bits

%%%%%%%%%%%%%%%%%%%%%%%%%%%%%%%%%%%%%%%%%%%%%%%%%%%%%%
\subsection{Resistance to Conventional Machine Learning Attacks}
\label{sec2.C}

%Now, we are shifting gear and starting to look at how the interface 
%can possibly thwart machine learning attacks.

Since the interface randomly chooses $n$ bits out of the $(n\!+\!m)$ input bits and feed the $n$ chosen bits to the $n$ stages of a PUF instance as challenge bits, attackers have no knowledge about which $n$ bits are used and which input bits are not used, leading to an obfuscation of the challenge. However, a trusted server of the PUF %, say a securely protected server partner of the PUF, 
knows which of the input bits are challenge bits and will be able to generate the response using a PUF model stored in the server to verify the PUF. 

But how many ghost bits are needed and how the ghost bits are
distributed among the input bits in order to thwart attacks
%interfaced APUFs
are important questions. That means, we need to identify 
\begin{itemize}
    \item[(i)] a set of interfaces, each member of the set have high potential to secure APUFs against machine learning attacks.
\end{itemize}
However, it is not enough to find a set consisting of a small number of  such interfaces. What we need is a set with huge number of PUF-securing interfaces, that is,
\begin{itemize}
    \item[(ii)] the set meeting the condition (i) must have exponentially many members.
\end{itemize}
If the set has a small cardinality, a brutal force exhaustive search of all interfaces in the set can identify which interface is implemented on a PUF instance. But if there are exponentially many interfaces meeting condition (i), the ghost bits of an interface behaves like a secret key. Thus, we need condition (ii) to make the interfaces secret-key-like.
Our study %along the direction of 
on issue~(i) has led to the following.

%%%%%%%%%%%%%%%%%%%%%%%%%%%%%%%%%%%%%%%%%%%%%%%%%%%%%
\vspace{1mm}
\begin{theorem}\label{thm1}
For an $n$-stage APUF equipped with an $m$-plus-bits challenge interface, if no pair of ghost bits are consecutive in the sense that $i_j + 1 < i_{j+1}$ for all $j=1, 2,\cdots,m\!-\!1$, 
then the response of the interfaced APUF as a function of the feature vector $\big(\phi(1), \phi(2), \cdots,\phi(n\!+\!m)\big)$ transformed from input $(b_1,b_2,\cdots,b_{n+m})$ according to
\begin{equation}
\label{thm1.1}
\phi(i) = (2b_i-1)(2b_{i+1}-1)\cdots(2b_{n+m}-1).
\end{equation}
is represented by a classification whose separation surface is defined by an $(n\!+\!m)$-variable polynomial of an order between $2m\!-\!2$ and $2m\!+\!1$.
%If there are ghost bits that are consecutive in the sense that $i_j + 1 = i_{j+1}$ for some $j\in\{1, 2,\cdots,m\!-\!1\}$, then the order of the polynomial will be lower.
\vspace{1mm}
\end{theorem}

%%%%%%%%%%%%%%%%%%%%%%%%%%%%%%%%%%%%%%%%%%%%

%%%%%%%%%%%%%%%%%%%%%%%%%%%%%%%%%%%%%%%%%%%%
\noindent{\em Proof} of Theorem~\ref{thm1}:  
Since $c_i$ is the bit fed to the $i$-th PUF stage, according to \cite{lim2005extracting}, the response of the interfaced APUF satisfies
\begin{equation}\label{thm1.2}
r = Sgn\Big( v(n) + \sum_{i=1}^{n} w(i)\phi^*(i)  \Big),
\end{equation} 
where $v$ and $w$'s are parameters quantifying gate delay differences at different stages, $Sgn(\cdot)$ is the sign function, and $\phi^*$'s are features transformed from challenge bits $c$'s according to 
\begin{equation}\label{thm1.3}
\phi^*(i) = (2c_i-1)(2c_{i+1}-1)\cdots(2c_n-1)
\end{equation}
for $i\!=\!1,\cdots,n$. However, attackers do not know $(c_1,\cdots,c_n)$, but know $(b_1,\cdots,b_{n+m})$, so they are not able to calculate $\phi^*$ but can calculate $\phi$ from $(b_1,b_2,\cdots,b_{n+m})$ according to (\ref{thm1.1}).  

Comparing (\ref{thm1.3}) and (\ref{thm1.1}) and utilizing (\ref{interface.3}), one can see that
\[
\label{Thm3.step4}
\left\{
\begin{aligned}
\phi(i)= &\,\phi^*(i)\,\prod_{j=1}^{m} (2b_{i_j}-1)
    & \text{ for } & i < i_{1}.\\
\phi(i)= &\,\phi^*(i\!-\!k)\! \prod_{j=k+1}^{m}(2b_{i_j}\!-\!1)\!
    &\!\text{for } & i_{k}\!<\!i\!<\!i_{k+1},\\
\phi(i)= &\,\phi^*(i\!-\!m) 	
    &\text{ for } & i > i_m,   
\end{aligned}
\right .
\]
for $k=1,2, \cdots, m-1$. 
Since $(2b_{i_j}-1)$ is either $1$ or $-1$, 
%for $j=1,2,\cdots, m$, 
from the equations above we can obtain
\begin{equation}
\label{Thm3.step5}
\left\{
\begin{aligned}
&\phi^*(i) ~ =\,  \phi(i)\, \prod_{j=1}^{m} (2b_{i_j}-1)
    & \text{ for } & i < i_{1}.\\
&\phi^*(i\!-\!k) =\,  \phi(i)\!
\prod_{j=k+1}^{m}\!
(2b_{i_j}\!-\!1) 
    & \text{ for } & i_{k}\!<\!i\!<\!i_{k+1},\\
&\phi^*(i\!-\!m) =\,  \phi(i) 	
    &\text{ for } & i > i_m,\,  
\end{aligned}
\right .
\end{equation}
for $k=1,2, \cdots, m-1$. 

When $i_m<n+m$, from the definition of $\phi$ in (\ref{thm1.1}) and utilizing the fact that $\phi(i)$ is either 1 or $-1$, one can obtain $(2b_{i_j}\!-\!1) = \phi(i_j)\,\phi(i_j+1)$ for all $i_j$, which, combining with (\ref{Thm3.step5}) to imply that
\begin{equation} \label{Thm3.step6}
\hspace*{-2.3mm}\left\{
\begin{aligned}
&\phi^*(i) ~ =\, \phi(i) \Big(\prod_{j=1}^{m} \phi(i_j)\,\phi(i_j+1)\Big)
    &\text{for } & i < i_{1},\\
&\phi^*(i\!-\!k)=\phi(i)\big(\!\!\!\prod_{j=k+1}^m \!\!\!\phi(i_j\!)\,\phi(i_j\!+\!1)\big)\!\!
    &\text{for } & i_{k}\!<\!i\!<\!i_{k+1},\\
&\phi^*(i\!-\!m)=\,  \phi(i) 	
    &\text{for } & i > i_m.\,  
\end{aligned}
\right .
\end{equation}
Thus, when $i_m\! <\! n\!+\!m$, Eqs.\,(\ref{Thm3.step6}) and (\ref{thm1.2}) combine to 
show that an $n$-stage APUF equipped with an $m$-plus-bits interface is a classification problem defined by a polynomial with $\phi$'s being the variables, and the order of the polynomial
is ($2m\!+\!1$) if $i_1>1$, but of order ($2m\!-\!1$) 
if $i_1=1$, since all these $\phi$'s on the right hand side of
(\ref{Thm3.step6}) are different when 
% $i_m<n+m$ and
any two ghost bits $i_{j}$ and $i_{j+1}$ are not consecutive.

%%%%%%%%%%%%%%%%%%%%%%%%%%%%%%%%%%%%%%%%%%
%Then (\ref{Thm3.step5}) and (\ref{Thm3.step2}) imply that an 
%$n$-stage arbiter PUF with an $m$-plus-bits interface is a 
%classification problem whose separation surface is expressible 
%by a polynomial with $\phi$'s being the variables.

When $i_m = n+m$, the term $(2b_{i_m}\!-1)$ in (\ref{Thm3.step5}) 
is equal to $\phi(n+m)$, not $\phi(i_m)\phi(i_m\!+\!1)$. So the 
representing polynomial will be of $(2m\!-\!2)$-th order or 
$2m$-th order depending on whether $i_1>1$ or not, one order 
lower than that for the case $i_m < n_m$ which corresponds
to Eq.\,(\ref{Thm3.step6}).
%
%Now, if two ghost bits at $i_{j'}$ and $i_{j'+1}$ are consecutive in the sense that 
%$i_{j'}\!+\!1 = i_{j'\!+\!1}$ for some $j'$, then 
%$(2b_{i_{j'}}\!-\!1) (2b_{i_{j'+1}}\!-\!1)$ is equal to
%\begin{equation} \label{ConsecutiveCase.1}
%%(2b_{i_{j'}}\!-\!1) (2b_{i_{j'+1}}\!-\!1) = 
%\phi(i_{j'})~\phi(i_{j'}+2),
%\end{equation}
%instead of
%\begin{equation} \label{ConsecutiveCase.2}
%%(2b_{i_{j'}}\!-\!1) (2b_{i_{j'\!+\!1}}\!-\!1) =
%    \phi(i_{j'})\,\phi(i_{j'}\!+\!1)\,
%    \phi(i_{j'\!+\!1})\,\phi(i_{j'\!+\!1}\!+\!1),
%\end{equation}
%since the product of the two middle terms in (\ref{ConsecutiveCase.2})
%is the same and therefore their product is equal to $1$ when $i_{j'}+1= i_{j'+1}$,
%and under this condition (\ref{ConsecutiveCase.2}) becomes (\ref{ConsecutiveCase.1}). 
%Then, for all $i < i_{j'}$, $\phi^*(i)$ is a polynomial %of $\phi$ 
%two orders lower than in the case that $i_{j'}$ and $i_{j'+1}$ are not consecutive. 
This completes the proof of the theorem.\hfill
\qedsymbol
%%%%%%%%%%%%%%%%%%%%%%%%%%%%%%%%%%%%%%%%%%%%%%%%%%%%%%%%%%

%\begin{itemize}
%    \item[] \hspace{-6mm}{\em $\bullet$~ Discussion of Issue (i)}
%\end{itemize}
%\noindent{\em $\,\bullet$~ Discussion on Issue (i)}\\

%In general, classification problems with highly nonlinear separation surfaces are more difficult to be machine learned than classifications with less nonlinear separation surfaces, and surfaces defined by higher-order multi-variate polynomials are more nonlinear. Thus, 
Theorem \ref{thm1} has addressed issue (i) that was raised in the paragraph before the theorem, that is, the the interfaced APUFs are potentially  % though not necessarily, 
secure against machine learning attacks when sufficient many ghost bits are used. 
%indicated by the order of the representing polynomials of the
%classification's separation surface. 
For instance, when $m\ge 20$, the representing polynomial of the classification separation surface is of 38-th order or higher.

%Theorem \ref{thm1} is about arbiter PUFs, and the interface can obviously be used for other PUFs. If the interface is applied to an $n$-stage $k$-XOR PUF either for the whole PUF or for individual component APUFs of the XOR PUF with different component APUFs equipped with an interface with different $G_m$'s, then it can be seen from Theorem~\ref{thm1} that the interfaced $k$-XOR PUF has a representing polynomial of an order around $2\,m\,k$ if no pair of ghost bits of an interface are consecutive.

Theorem \ref{thm1} indicates high potential for these interfaces to secure APUFs against machine learning attacks when the interfaces have a large number of ghost bits. But exactly how large the value of the parameter $m$ can lead to secure APUFs have to be experimentally determined. Thus, in Sec.\,\ref{sec4_Experiment}, we will experimentally examine how well interfaced PUFs perform against attacks with different values for interface parameters.

%\begin{itemize}
%    \item[] \vspace{-1mm}
%    \item[] \hspace{-6mm}{\em $\bullet$~ Discussion of Issue (ii)}
%\end{itemize}
%\noindent{\em $\,\bullet$~ Discussion on Issue (ii)}\\

For issue (ii), it can be verified that the number of sets $G_m$'s satisfying $n/3\le m\le n/2$ and $i_j+1 < i_{j+1}$ for all $j=1,2, \cdots, m-1$ grows exponentially with $n$. 
This can be seen by choosing $m=n/2$ and $i_1$ from $\{1,2\}$, $i_2$ from $\{4,5\}$, $\cdots$, $i_j$ from $\{3j-2,3j-1\}$ for $j=1,2,\cdots, m$, resulting in $2^{m} = 2^{n/2}$ possible sets $G_m$'s whose associated polynomials are of order of $2(m\!-\!2)$ or higher. 
%This can be seen that for $m=n/3$, one can have three choices for $i_1$ from $\{1,2,3\}$, three choices for $i_2$ from $\{5,6,7\}$, $\cdots$, and three choices for $i_j$ from $\{4j\!-\!3,4j\!-\!2,4j\!-\!1\}$ for $j\!=\!1,2,\cdots, m$, resulting in $3^{m} > 2^{1.5\,m}= 2^{n/2}$ possible sets $G_m$'s whose associated polynomials are of order of $2(m\!-\!2)$ or higher. 
So there are exponentially $m$-plus-bits interfaces which will turn an APUF into a classification problem defined by high-order polynomials.

\section{The Authentication Protocol}\label{sec3_authentication}

Communications between an IoT device and a server may involve commands from the server to the device or sensor data from the device to the server. It is helpful, and sometimes necessary, for the device to authenticate the source of a command it receives and for the sever to verify the source of the sensor data. Hence, we are focused on mutual authentication.

\subsection{Technical Assumptions}\label{sec3.Assumptions}

Our protocol is to authenticate the communications between a resource-constrained PUF-embedded device and a securely protected resource-rich server. We assume that 
\begin{itemize}
    \item[1.] the communication channel between the server and PUF is publicly accessible and all communication data, including CRPs, can be seen by third parties;\vspace{0.5mm} 
    \item[2.] intermediate transient operation results on the device, including data temporally stored in registers, are not accessible by any party other than the device itself;\vspace{0.5mm} %(that is, side-channel attacks are not considered in this paper), 
    \item[3.] the device has a small nonvolatile memory (NVM) with a capacity far inadequate for storing all challenges used or to be used in the operation life of the  protocol, and data on the NVM are openly accessible but can be altered only by the device (or intrusions are detectable and will lead to suspension of device operation); and
    \item[4.] the PUF physical variations that determine the PUF responses and the PUF internal wiring %that are part of the hardware secrete key of the PUF and 
    are inaccessible to any third-party.\vspace{0.5mm}
\end{itemize} 

One of the goals of our authentication protocol is to prevent successful modeling attacks of the PUF, including both conventional and reliability-based machine learning (ML) attacks. A reliability-based ML attack requires that each challenge be evaluated by the PUF multiple times and the CRPs be accessible by attackers.
Conventional ML attacks are a more prevalent type of non-physical threats to strong PUFs. Suck attacks refer to those where attackers accumulate CRPs passively, like eavesdropping on the communications between the PUF and its trusted party. The accumulated CRPs are then used by the attackers to build machine learning models that can predict future responses of the PUFs after the models are trained with sufficient CRPs. Passive CRP accumulation becomes a choice for attackers when the PUF, under the control of its managing protocol, responds to an input challenge only after some security hurdle is cleared, e.g. after authentication of the source of some message received by the device, %from PUF's trusted server 
as in mutual authentication protocols.

%%%%%%%%%%%%%%%%%%%%%%%%%%%%%%%%%%%%%%%%%%%%%%%%%%
\subsection{Technical Requirements}

\subsubsection{Requirements to Fight Reliability-based ML Attacks}$~$

%Before presenting the protocols for authenticating the communications in and out of the PUF-embedded device, we  
Given the assumptions, we starts with discussing technical requirements on the protocol. %to make the protocols resistant to both conventional and reliability-based modeling attacks. 
These requirements are to be implemented on legitimate operations of the protocol with a goal of thwarting illegitimate activities including conventional and reliability-based modeling attacks.

Since reliability-based ML attacks require (i) repeated evaluation of same challenges and (ii) access to these CRPs by attackers, denying one of the two can prevent such attacks. Hiding or obfuscating either the challenges or the responses can prevent third-party accesses to CRPs, and repeated evaluations of challenges can be prevented if every challenge fed to the device is a fresh challenge never used before. Challenge freshness has been adopted by many protocols and will be part of our strategy to prevent reliability-based machine learning attacks.  %Thus, one of the directly implementable technical requirements we decide to impose on our protocol is the following.
The following challenge freshness is the start of a technical requirement that will be presented later.%we are developing.

\vspace{1mm}
\begin{itemize}
    \item[R0.]  Every challenge used in legitimate authentications must be a fresh challenge never used in earlier authentications.\vspace{1mm}
\end{itemize}

We wish to comment that the resource-constraint as specified in Assumption 3 is a factor for R0 in order to fight reliability-based machine learning attacks. This can be seen from a case of a resource-rich device with a large secure NVM. For such a device, during the enrollment event before deployment of the device, in a secure environment all challenges needed for the entire operational life are stored into the device's secure NVM and also into server's secure storage. Then, in each authentication, the device selects the first challenge in the challenge database, feeds it to the PUF, send the responses to prove itself, and then deletes the challenge after the authentication. The server does the same to pick the first challenge from its own database and deletes it after the authentication. Since only responses but no challenges are transmitted through the communication channel, attackers have access only to responses but no access to CRPs, and hence authentications can function effectively and securely even if there are some repeated challenges. Thus, R0 would not be necessary for fighting reliability-based machine learning attacks without Assumption~3.

If attackers are restricted to collect CRPs only by passively listening to legitimate communications in and out of the PUF, R0 will be sufficient for preventing reliability-based ML attacks. But R0 stipulates only that every challenge used in legitimate authentications be fresh, but does not specify solution for limiting attackers to only passive CRP collection. Let us look at a scenario where a mutual authentication protocol is designed to have met R0 in a way that a PUF's trusted server supplies a fresh challenge together with a subsequence of responses %the server has generated using a soft model of the PUF for authenticating
to authenticate the server itself to the PUF. 
Then, attackers can collect legitimate CRPs passively, but can illegitimately supply to the PUF a used challenge together with the response subsequence to masquerade as the server. 
%Requirement R0 is not sufficient for preventing reliability-based ML attacks, unless attackers can collect CRPs only passively by listening to communications between the PUF and its trusted partners. %since ensuring freshness of challenges used in ligtimate authentications could still leave room for attackers to supply used challenges to the PUF. 

The scenario shows that, in addition to R0, we must also fight replay attacks,  where a replay attack is one in which some challenges already used by the protocol in earlier authentications are re-supplied to the device by an attacker, raising the risk of allowing adversaries to issue commands to the device.  The replay attack is itself a threat and also a possible enabler of reliability-based attacks.  Thus, we impose the following requirement for the protocol.

\begin{itemize}
\item[R1.] For the challenge used in each legitimate authentication, 
\begin{itemize}
    \item[a.] the device supplies some bits of the challenge, and
    \item[b.] the device-supplied part of any challenge is fresh.\vspace{0.5mm}
\end{itemize}
%the device must supply some bits of the challenge, and the device-supplied part of the challenge must be fresh. 
%, and the device-supplied challenge bits are random with an equal probability of being 0 or 1.
\end{itemize}

The R1.b is a stronger requirement than R0, since a fresh challenge (satisfying R0) can still have device-supplied part of the challenge not fresh. An implication of R1 is that the number of device-supplied challenge bits needs to be large enough, because it is not possible to maintain the freshness of a short device-supplied part of the challenge for the protocol designed to handle a large number of authentications during its operation life. For instance, if only 10 bits are device-supplied in a protocol that is expected to execute over thousands of authentications, challenge freshness cannot be maintained after executing 1024 authentications. 
%The validity of Requirement R2 can be viewed from the following scenario. Suppose that in a mutual authentication protocol, the device does not supply a single bit of the challenge.  The server generates the entire challenge and sends it to the device together with a subsequence of responses generated using a soft model by the server for authenticating the server itself to the device. 

Requirement R1 is to prevent replay attacks like the one discussed in the paragraph preceding R1. For a device without a large NVM to store all used challenges (Assumption~3), the device has no way to keep track of used challenges.  Without R1, even R0 cannot prevent an attacker from re-sending used challenges and response subsequences collected from the open communication channel (Assumption~1). Thus, Requirement 1 is essential for fighting replay attacks under the assumptions given in Sec.\,\ref{sec3.Assumptions}. A rigorous statement on the security of the Requirement R1 against replay attacks is given below.
%replay attacks become possible if a protocol is not carefully designed.

%%%%%%%%%%%%%%%%%%%%%%%%%%%%%%%%%%%%%%%%%%%%%
\vspace{1mm}
\begin{theorem}\label{thm2}
Under the assumptions given in Sec.\,\ref{sec3.Assumptions}, Requirement R1 is a sufficient and necessary condition for a PUF-based mutual authentication protocol to defeat replay attacks.\vspace{1mm}
\end{theorem}
\noindent{\em Proof} of Theorem~\ref{thm2}: 
For a protocol meeting Requirement R1, any replay attack has to send non-device-supplied part of a used challenge and a response subsequence for the used challenge. But the device-supplied part of any used challenge will not be supplied by the device due to R1.b, and hence any replay attack cannot be successful, which establishes R1 as a sufficient condition. 

To prove R1 as a necessary condition, we will examine two cases, where in case one R1.a is violated, and in case two, R1.a holds but R1.b is invalid. In case one, the scenario given in the paragraph that is two paragraphs ahead of Requirement R1 shows that there exist room for replay attacks when R1.a does not hold.

In case two where R1.b does not hold, the device will not reject an authentication request if the device-supplied part of a challenge is the same as the device-supplied part of an earlier legitimate challenge. Hence, an attacker can send a non-device-supplied part of the earlier challenge together with the response subsequence, and the device will likely accept the authentication request since the device has no way to keep track of all not-device-supplied parts of challenges without a large NVM to store all of them (Assumption~3), leading to the success of a replay attack.

We wish to comment that when R1.b is violated but R0 holds (meaning the not-device-supplied part of challenges used in legitimate operations are fresh), the attack described in case two of the necessary condition will remain successful, showing that R1.a plus R0 is insufficient to fight replay attacks.

Thus, under either case one or case two, relay attacks become possible, meaning that Requirement R1 1 is a necessary condition. This completes the proof of the theorem.\hfill
\qedsymbol

%%%%%%%%%%%%%%%%%%%%%%%%%%%%%%%%%%%%%%%%%%%%%%%%%%%%%%
\subsubsection{Requirements to Fight Conventional ML Attacks}$~$

%Besides reliability-based machine learning attacks, a more prevalent type of no-physical threats to strong PUFs is the conventional machine learning attacks. Suck attacks refer to those where attackers accumulate CRPs passively, like eavesdropping on the communications between the PUF and its trusted party. The accumulated CRPs are then used by the attackers to build machine learning models that can predict future responses of the PUFs accurately after the models are trained with sufficient CRPs. Passive CRP accumulation becomes a choice for attackers when the PUF, under the control of its managing protocol, responds to an input challenge only after some security hurdle is cleared, e.g. after the authentication of the source of the non-self-generated part of the challenge as in mutual authentication protocols.

%If a modeling-resistant PUF design is used for authentication, the authentication protocol, without considering the PUF which could be sophisticated, can be designed with high simplicity and lightweightness by relying on the PUF to deliver security against modeling attacks. 

%Every protocol must be examined for its potential security vulnerabilities before being deployed. It is our belief that simplistic protocols take less effort for protocol developers to identify all possible vulnerabilities, and hence are less likely to have undiscovered ones than more complex protocols.

To fight conventional modeling attacks, we plan to use a modeling-attack-resistant PUF. Our philosophy is that if the PUF used in an authentication protocol is resistant to modeling attacks, the protocol itself, without considering the PUF which could be sophisticated, can be designed with high simplicity and low resource overhead by relying on the PUF to deliver security against modeling attacks. Simplistic protocols would take people less effort to identify their possible vulnerabilities
%are easier for vulnerability identification 
and also incur low implementation overhead for the PUF-excluded part of the protocol. Protocol II in \cite{Yu2016lockdown} employs this approach with the use of a component-differential-challenged XOR PUFs. %which, at the time this paper is under preparation, is secure against modeling attacks \cite{tobisch2015scaling,Li2022New} for such PUFs with 7 or more 64-bit component APUFs. 

We are considering to use the highly lightweight interfaced APUF proposed in Sec\,\ref{sec2}. The challenge obfuscation interface requires a condition to protect the obfuscation from being exposed. If the interfaced PUF is openly accessible and responds to any challenge, an attacker can choose challenges of particular bit patterns and the responses of the interfaced PUF could reveal the mapping function of the obfuscation. An exposed obfuscation mapping function will enable easy machine learning attacks of the interfaced APUF.

%The bit patterns of such chosen-challenge attacks involve no repeated challenges (hence Requirement R1 is irrelevant for helping or fighting such attacks), and are highly unlikely to appear in a challenge of random bits. 

Requirement R1 
%prevents attackers from choosing the bit pattern for the device-supplied part of the challenge, 
helps reduce the exposure of the obfuscation by eliminating the opportunity for attackers to choose the %bit pattern of 
device-supplied challenge bits. But challenge bits that are not supplied by the device could possibly leave room for attackers if the protocol is not carefully designed. From the discussion near the end of Sec.\,\ref{sec2.B}, the likelihood to defeat chosen-challenge attacks also depends on the number of bits which are random with an equal probability of being 0 or 1. 

Even if device-supplied challenge bits are random and fresh, as long as the number of such challenge bits is not large enough, there might be chances for chosen-challenge attacks. The number of device-supplied challenge bits needed for fighting chosen-challenge attacks could be much larger than that required by R1 for fighting replay attacks.

For instance, if in a mutual authentication protocol designed for executing less than 4 billion authentications with the use of a 64-bit PUF, 32 challenge bits are device-supplied and made fresh by using a 32-bit maximum-length LFSR, and 32 remaining challenge bits together with 32 response bits are supplied by the server, and the device  accepts an authentication request after the first 32 response bits generated by the device for the challenge match the received 32 response bits, within the protocol's error threshold. Then, upon an authentication request from an attacker, the device may generate 32 device-supplied challenge bits that are the same as the device-supplied part of an used legitimate challenge on exactly 31 bits (hence still fresh due to difference on 1 bit), the attacker can reply with the 32 server-supplied bits of the used challenge and the 32 response bits from the earlier legitimate authentication. If the authentication request is accepted by the device, the mis-matched challenge bit on the device-supplied part has a high probability to be a ghost bit, and if multiple attacks of the this type happen to have the same mis-matched challenge bit on the device-supplied part, this bit is almost certainly a ghost bit. Of course, such an attack could take long time since it may take many authentication requests to have one pair of requests that have the two device-supplied challenge parts identical on 31 bits, and take even substantially more authentication requests to have multiple requests that have the same mis-matched bit. But after all, chosen-challenge attacks become possible under this protocol.
%if the protocol is designed to allow the device to respond to any non-device-supplied challenge bits. 

Thus, our strategy to eliminate opportunities for chosen-challenge attacks is that in every legitimate authentication,

\begin{itemize}
    \item[R2.] all challenge bits are random with an equal probability of being 0 or 1, and the device-supplied part of the challenge is long enough to make chosen-challenge attacks impossible.
\end{itemize}

%\begin{itemize}
%    \item[R2.] if the device does not have database for storing all challenges that are to be used in all authentication operations, mutual authentication is necessary.
%\end{itemize}

%Keeping a challenge database on the device can meet requirement R2 by picking the first challenge from the database, using it for authentication, and then deleting it from the database. This approach was used by \cite{Lee2013Mutual}. But with assumption of limited resources and no challenge database on device, requirement R2 means that part of the challenge must come from the device. In the lockdown protocol II \cite{Yu2016lockdown}, part of the master challenge is generated on the device using a TRNG.

%A natural strategy to ..., and this is what several existing protocols, including the highly lightweight lockdown protocol \cite{Yu2016lockdown}, have done.

\subsection{The Proposed Mutual Authentication Protocol}

In the following, we presents a protocol that performs mutual authentication of communications between a device and a server, where an interfaced APUF has been embedded on the device and soft model of the PUF is available on the server.  The protocol consists of an  enrollment phase and an authentication phase, where the enrollment is a one-time event. We use the terminology an authentication session to denote a sequence of authentications in a duration in which the device is in continuous operation without being powered off, and an authentication session consists of one or multiple authentications.% with different authentications in a session authenticating the sources of different messages.  
We also use the terminology master challenge to denote the seed fed to a pseudo-random generator (implemented by an LFSR) to generate a sequence of derived challenges to be evaluated by the PUF.

%The protocol is to authenticate  communications initiated either by the device or by the server. In a server-initiated communication operation, the server  sends to the device a message which contain metadata to enable the device to authenticate if the message is from the server, and in every message the device sends to the server, 

Before presenting the enrollment and authentication, we list the device-side protocol-supporting architectural composition. 
\begin{itemize}
\item An $n$-stage challenge obfuscation interfaced APUF with $(n\!+\!m)$ input bits, where$n\ge64$ and $m$ is an integer between $n/3$ and $n/2$;
%\item  an $(n\!+\!m)$-bit register R1 to hold the challenge;
\item an $(n\!+\!m)$-bit maximal-length Linear Feedback Shift Register (LFSR) as a pseudo-random number generator (PRNG);
\item an $n/2$-bit register R2 to hold $n/2$ response bits received from the server;
\item an $n$-bit register R3 to store $n$ responses generated by the PUF from a master challenge; 
\item a comparator for comparing if two $n/2$-bit vectors have a Hamming distance below an authentication acceptance threshold; and
\item a small NVM  %of at least $(n\!+\!m)$ bits to store an $(n\!+\!m)$-bit challenge for the first authentication of the next session. The $(n\!+\!m)$-bit vector on the NVM, denoted by $C_{nvm}$, is the value-changing ID of the device and is made publicly accessible.
to store an $(n+m)$-bit ID of the device, which will be updated during each authentication. This ID is denoted by $C_{\rm nvm}$ and is made publicly accessible. 
\end{itemize}

The server has the following units and variables for the authentication protocol. 
\begin{itemize}
\item An $(n\!+\!m)$-bit LFSR which behaves exactly the same as the LFSR on the device; 
\item a soft model of the PUF on device, which is to be trained during the enrollment phase and assumed to generate, with a high probability of correctness, the response of the PUF for a challenge given to the PUF; 
\item a comparator for comparing if two $n/2$-bit vectors have a Hamming distance below an authentication acceptance threshold; and
\item an $(m+n)$-bit variable ${\rm ID_{device}}$ that stores the changing ID of the device.  It is the server counterpart of $C_{\rm nvm}$ on the device, will be initialized during enrollment, and updated in each authentication. 
\end{itemize}

Now, we are ready to present the authentication protocol, and we start with the enrollment.
\vspace{2mm}

\noindent{\em The One-Time Enrollment Phase}
\begin{itemize}
\item[E1.] Before the PUF is embedded into the device, the server collects a number of CRPs from the challenge obfuscated APUF to train a soft model to a satisfactory accuracy level. With the challenge obfuscation known to the server, the model training needs only a small number of CRPs, but the obfuscation is assumed to be unknown to any third party.\vspace{0.5mm}
\item[E2.] After completion of the training of the soft PUF model, the PUF is embedded into the device.\vspace{0.5mm}
\item[E3.] Assign an $(n+m)$-bit initial id to the device, store this initial id in $C_{\rm nvm}$ on the device NVM, and also store it in an $(n+m)$-bit ${\rm ID_{device}}$ on the server. This device id will be updated in authentications, so it is called the initial device id during enrollment.
\end{itemize}

%The protocol is to authenticate  communications initiated either by the device or by the server. In a server-initiated communication operation, the server  sends to the device a message which contain metadata to enable the device to authenticate if the message is from the server, and in every message the device sends to the server,

We choose to let the server initiate an authentication, as is done in several PUF-based authentication protocols \cite{Yu2016lockdown,Qureshi2019PUF_RLA,Gu2021Modeling}. %The protocol is to authenticate  communications initiated either by the device or by the server. 
In a server-initiated communication operation, the server sends to the device a message containing metadata to authenticate the server as the source of the message, and also enable the device to prove itself as the source of a response message sent by the device to the server.\vspace{3mm}

\noindent{\em The Authentication Phase}\vspace{0.5mm}

%\noindent{\bf Server}
\begin{itemize}
\item[]\hspace{-7mm}Server
    \begin{itemize} 
    \item[A1.] The server loads ${\rm ID_{device}}$ into the LFSR as the seed for the LFSR to generate $n$ challenges of $(n\!+\!m)$ bits each, feeds the $n$ challenges to the soft PUF model to get $n$ responses.
    \item[A2.] Then the server sends a vector of $(n/2)$ bits to the device, with the $(n/2)$ bits from the first half of the $n$ responses generated by the soft PUF model.\vspace{1mm}
    \end{itemize}

%\noindent{\bf Device}
\item[]\hspace{-7mm}Device 
    \begin{itemize}
    \item[A3.] The device receives the $(n/2)$-bit vector from the sever and stores them into R2, and then loads $C_{\rm nvm}$ into the LFSR as the master challenge for this authentication. 
    \item[A4.] The device goes through $n$ iterations as in the following:
    starting with the master challenge as the first challenge, in each iteration the bit vector in the LFSR is used as the challenge to the PUF to get 1-bit PUF response and store it into R3, and then the LFSR generate the next $(n\!+\!m)$-bit vector.
    
    (Comment: After the $n$ iterations, all $n$ bits of PUF responses are generated and stored into R3, and what remains in the LFSR is a bit vector that has not been fed to the PUF)
    \item[A5.] The device compares R2 with the first $n/2$ bits of R3.  
    \item[A6.]
        \begin{itemize}
        \item[if]
        Hamming distance of comparison is larger than threshold, the device aborts the authentication; 
        \item[]\hspace{-4mm}else
        
        the device authenticates the server, sends the last $n/2$ bits of R3 to the server, and  stores the bit vector in the LFSR into $C_{\rm nvm}$ on the NVM.\vspace{1mm}
        \end{itemize}
        
    \end{itemize}
\item[]\hspace{-7mm}Server
%\noindent{\bf Server}
    \begin{itemize}
    \item[A7.] The server receives $n/2$ bits from the device and compares them with the second half of the $n$ response bits out of the PUF soft model, 
    \item[A8.]
        \begin{itemize}
        \item[if] the Hamming distance is larger than the threshold,
        the server rejects the received message;
        \item[]\hspace{-4mm}else
        
        the server authenticates the device as the source of the message, and then stores the current bit vector in the LFSR into ${\rm ID_{device}}$.
        \end{itemize} 
   \end{itemize}
\item[]\hspace{-6mm}{\em End of an authentication}

\end{itemize}

%%%%%%%%%%%%%%%%%%%%%%%%%%%%%% End of Sec3.tex %%%%%%%%%%%%%%%%%%%%%%%%%%%%%%

\section{Discussions on the Lightweightness of the Protocol}

In this section, we discuss the device-side hardware, operation and communication overheads of the protocol proposed in the preceding section, and compare the overheads with those of existing PUF-based authentication protocols.

\subsection{Communication and Operation Overheads for the Device}
\subsubsection{Communication Overhead} $~$

One difference with most, if not all, PUF-based protocols is that our protocol does not use a permanent-value device ID, but a value-changing device ID which is also used as the master challenge. %when the authentication is the first one in a session. 
By sharing this ID with the server during enrollment and updating it with an LFSR which is known to the server (and also assumed to be known to attackers), we have eliminated the need to communicate the challenges between the device and the server. Actually, all communication metadata used for one authentication event are $(n/2)$ bits from the server to the device in Steps A2-A3 and $(n/2)$ bits from the device to the server in Steps A6-A7. 

On the amount of communication data and number of PUF evaluations used in protocols, Idriss et al.~\cite{Idriss2021Lightweight} compared four PUF-based authentication protocols \cite{Majzoobi2012Slender,Yu2016lockdown,Gu2021Modeling,Idriss2021Lightweight}, and the protocol of Gu et al.~\cite{Gu2021Modeling} has the lowest amount of communication data among the four protocols and also the lowest number of PUF evaluations. %(evaluations of the False PUF plus the evaluations of the Genuine PUF).  
Upon examining the protocol \cite{Gu2021Modeling}, we found that the amount of the communication metadata of our protocol is about half of that of the protocol \cite{Gu2021Modeling}. Specifically, the protocol \cite{Gu2021Modeling} sends a half-length challenge from server to device and sends another half-length challenge concatenated with device ID from device to server, which, at this point, has already passed amount of the communication data of our protocol (due to sending the device ID), and then executes a pair of send-receive operations between the device and server.
We also found that the number of PUF evaluation of our protocol might be slightly lower than that of the protocol of Gu et al.~\cite{Gu2021Modeling}, which performs evaluations of the False PUF and the Genuine PUF. The numbers of response bits generated from the evaluations of the False PUF and the Genuine PUF are not known, but understanding of the protocol \cite{Gu2021Modeling} leads to an estimate of $n/2$ response bits out of evaluations of the False PUF and $n$ response bits out of evaluations of the Genuine PUF, amounting to a total of $3n/2$ bits. As a comparison, our protocol evaluates the PUF to get $n$ response bits.

\vspace{1mm}
\subsubsection{Operation Overhead} $~$

For device-side operation complexity, our protocol is operationally simplistic, and the major operations executed on the device include generation of $n$ challenges for the PUF, $n$ evaluations of the PUF, and a comparison of two $(n/2)$-bit vectors.  Mutual authentication protocols usually need to perform these operations on the device if the device receives only one challenge to produce $n$ responses.  The only protocols under which the device does not need to generate $n$ challenges are protocols that let the device receive $n$ challenges, but at the communication overhead of receiving $n$ challenges in addition to the remaining operations our protocol executes ($n$ evaluations of the PUF and a comparison of two bit-vectors of about $(n/2)$ bits).

%%%%%%%%%%%%%%%%%%%%%%%%%%%%%%%%%%%%%%%%%%%%%%%%%%%%%%
\subsection{Hardware Overhead for the Device}
%Regarding hardware overhead of the protocol, 
\subsubsection{PUF-Excluded Hardware Overhead} $~$

The supporting architecture of our protocol consists of an $(n\!+\!m)$-bit LFSR (a PRNG), two other registers (one $n/2$ bits and the other $n$ bits), and a comparator of two $(n/2)$-bit vectors, and an PUF. To the best of our knowledge, all mutual authentication protocols that receive only one master challenge for generating $n$ bits of responses need similar hardware units for the device. 

Some protocols \cite{Zhang2021Set_based,Chen2022Novel,Idriss2021Lightweight} do not use LFSR for the device but either let the device receive $n$ challenges for generating $n$ bits of responses, or use a TRNG to generate challenges which need to send to the server. One protocol \cite{Gao2016Obfuscated} does not explicitly mention LFSR, and from the description of the protocol an LFSR or a bit-vectors generator is necessary since one challenge is sent to the device to generate a multiple bits of PUF responses. 

In addition, most of existing protocols \cite{Majzoobi2012Slender,Yu2014Noise,Yu2016lockdown,Gao2016Obfuscated,Zhang2021Set_based,Gu2021Modeling,Idriss2021Lightweight,Chen2022Novel} need a TRNG for the device, and to the best of our knowledge only one protocol \cite{Zalivaka2019Reliable} uses an LFSR but does not need a device-side TRNG. The protocols \cite{Zhang2021Set_based,Chen2022Novel,Idriss2021Lightweight} do not have device-side LFSR (or PRNG) but have a TRNG on the device. As shown in \cite{Chen2022Novel}, a TRNG implemented on FPGA has about twice hardware overhead as that of an APUF.  We estimate that an $(n\!+\!m)$-bit LFSR, for $m\le n/2$, has an area overhead below 0.9 times of that of an $n$-bit APUF, meaning that our LFSR probably incurs a lower hardware overhead than a TRNG. Thus, before considering the PUF, our protocol is among the protocols with the lowest device-side hardware overhead among all protocols we are aware of \cite{Majzoobi2012Slender,Yu2014Noise,Yu2016lockdown,Gao2016Obfuscated,Zhang2021Set_based,Gu2021Modeling,Idriss2021Lightweight,Chen2022Novel,Zalivaka2019Reliable}.  
%We do agree that depending on particular devices, some of these units can be shared with other operations executed by the device, as in the work \cite{Che2017Privacy} where , so the actual area overhead is not easy to estimate accurately.

\vspace{1mm}
\subsubsection{Overhead of the PUF} $~$

On the overhead of the PUF, our protocol uses one APUF with a zero-transistor obfuscation interface. Several of the highly secure protocols either use multiple APUFs 
%(as in \cite{Gao2016Obfuscated,Gu2021Modeling}) 
or use a PUF with multiple APUFs as components. 
%(as in \cite{Majzoobi2012Slender,Yu2014Noise,Yu2016lockdown}), 
%so our protocol has a lower PUF area overhead than these protocols.
For instances, the protocol of Gao et al.~\cite{Gao2016Obfuscated} also uses the highly lightweight APUF, but uses three APUF instances to resolve collisions resulting from the intersection of a randomly selected bit groups from two pre-given bit groups. The highly simplistic lockdown technique-based protocol II \cite{Yu2016lockdown} uses a CDC XPUF, which, at the time the paper is under preparation, has to have more than 6 component APUFs \cite{tobisch2015scaling,Li2022New} to stay secure against the logistic regression-based modeling attack \cite{ruhrmair2010modeling}. The communication-efficient and operation-efficient protocol of Gu et al.~\cite{Gu2021Modeling} uses two PUFs. So our protocol has a lower PUF area overhead than these protocols.

There are also protocols \cite{Zalivaka2019Reliable,Zhang2021Set_based,Idriss2021Lightweight,Chen2022Novel} that use only one APUF. The protocol of Zalivaka et al.~\cite{Zalivaka2019Reliable} uses an APUF with its input challenge obfuscated by an MISR unit consisting of a D flip-flop, two XOR gates and two AND gates for each challenge bit, which we estimated to have about 1.5 times area overhead of that of an APUF. 

The two protocols A and B of Chen et al.~\cite{Chen2022Novel} use a device-side TRNG and other hardware units but no LFSR, and each protocol sums up a device-side overhead much higher than that of an LFSR. 
Specifically, the one-sided authentication protocol A of Chen et al.~\cite{Chen2022Novel} employs TRNG-based random shuffle of response groups to obfuscate the response, where the TRNG and the shuffle are additional overhead besides the APUF with the shuffle estimated by us to have an area overhead of at least half of that of an APUF and the TRNG shown in \cite{Chen2022Novel} to have a higher FPGA implementation overhead than the APUF. The mutual authentication protocol B of Chen et al.~\cite{Chen2022Novel} uses an APUF, a TRNG, and a major additional hardware overhead which was shown by themselves \cite{Chen2022Novel} to have an FPGA implementation close to that of the APUF. Both protocols A and B in ~\cite{Chen2022Novel} do not need LFSR for the device but at the cost of receiving $n=k\cdot q$ challenges for generating $n$ responses.
%with hash or random generated points for polynomials 
The one-sided authentication protocol of Zhang and Shen \cite{Zhang2021Set_based} has a low device-side hardware overhead with a PUF, a TRNG, a small NVM, and no LFSR, slightly higher than the overhead of our protocol. But it is unknown if it is possible to modify the protocol of Zhang and Shen \cite{Zhang2021Set_based} into a mutual authentication protocol at an affordable additional hardware/operational overhead.  %, but at a communication cost of receiving $n$ challenges for generating $n$ responses.
The mutual authentication protocol of Idriss et al.~\cite{Idriss2021Lightweight} introduces a new technique to obfuscate both the challenge to and the response from an single APUF, and its major device hardware overhead besides the APUF is a TRNG (LFSR is not needed).  Thus, compared with these five single-APUF-based protocols, our protocol also has an advantage in hardware overhead.

%We do agree that depending on devices, some hardware units can be shared with other operations executed in the device, as in the work \cite{Che2017Privacy} where data paths in some hardware units, like the hardware implemented AES, are used to generate delay differences that are used to produce circuit-dependent responses, so the actual hardware overhead for devices that employ unconventional PUFs like the one in \cite{Che2017Privacy} is difficult to compare. 

\section{Security of the Protocol against Replay and Reliability-based Modeling Attacks}

Letting the device take all challenge bits from the device and using a maximal-length LFSR as a PRNG to generate challenges using the initial value of the device ID as the seed, the protocol practically guarantees the freshness of the device-supplied part of every challenge used in legitimate authentications due to the extremely long cycle of $2^{n+m}$ for  a maximal-length LFSR. This means that the protocol satisfies Requirement R1, which prevents replay attacks and reliability attacks as revealed by  Theorem~\ref{thm2}.

%With both Requirements 1 and 2 met, replay attacks and reliability attacks are prevented,  since all challenges fed to the PUF are coming from the device and are fresh, leaving no opportunity for any third party to evaluate the PUF multiple times with same challenges or supply a response to a used challenge.

%%%%%%%%%%%%%%%%%%%%%%%%%%%%%%%%%%%%%%%%%%%%%%%%%%%%%%
\section{Security of the Protocol against Conventional Modeling Attacks}

The security of the protocol against conventional modeling attacks obviously depends on the protocol's capability of defeating chosen-challenge attacks.% and (ii) the security of the interfaced PUF against machine learning attacks when the challenges collected by attackers are random. 
The use of a maximal-length LFSR means that every bit of the challenge is random with equal probability of being 0 or 1 within the cycle of the LFSR. In addition, by letting the device take all challenge bits from the device, the protocol meets Requirement R2 as long as the the number of challenge bits $2^{n+m}$ is large enough, and discussions near the end of Sec.\,\ref{sec2.B} show that $n=64$ and $m=20$ is sufficient. Thus, chosen-challenge attacks are prevented by the protocol.
%and then the security of the protocol will be determined by the security of the interfaced APUF against modeling attacks wit

According to Kerckhoff's principle, attackers are assumed to know the design of a security system but not the secret key. For this protocol, all responses are sent over open communication channels, and all challenges fed to the interfaced PUF are also publicly known since the initial value of the device ID is known publicly and the LFSR is assumed to be known publicly. The ``secret key'' of the PUF, as specified in Assumption~4, is made up of the physical variations that determine the PUF responses and the mapping that implements the challenge obfuscation.  Thus, all CRPs used in authentications are available to attackers, providing abundant data for training machine learning attack models. Thus, the security of the protocol against conventional modeling attacks is determined by the security of the interfaced APUF against modeling attacks when all CRPs are openly accessible CRPs. 

In general, classification problems with highly nonlinear separation surfaces are more difficult to be machine learned than ones with less nonlinear separation surfaces. Theorem~\ref{thm1} shows that the separation surface of an interfaced APUF is defined by a high-order multivariate polynomial when the number of ghost bits, $m$, is large. But exactly how large the value of $m$ can lead to secure APUFs have to be experimentally determined. In the following, we will experimentally examine how well interfaced APUFs perform against modeling attacks with different numbers of ghost bits.

%We believe that any randomly or pseudo-randomly generated sequence of challenges is not likely to have such bit patterns. In the protocol, the challenges fed to the PUF are generated by an LFSR on the device, which is a PRNG. Thus, the protocol meets Requirement 2 which is a condition for the PUF to stay secure against chosen-challenge attacks, and the resistance to chosen-challenge attacks is key to the resistance to machine learning attacks.  And if and how secure the PUF is against machine learning attacks  and how chosen-challenge attacks can be defeated with high probability will be discussed in depth in Sec.~\ref{sec4_PUF}.
% the security of the protocol against modeling attacks depends on the PUF, whose resistance to machine learning attacks

\subsection{Experimental Attack Study of Interfaced APUFs}\label{sec4_Experiment}

%Since the interfaced APUF is an essential piece of our lightweight modeling-attack-resistant authentication protocol presented in Sec.\,\,\ref{sec3_authentication}, we carry out machine learning based modeling attacks on the interfaced APUFs with different numbers of ghost bits. 

\subsubsection{The Modeling Attack Method Used in Attack Study}\label{sec_NN_AttackMethods}
$~$

%\subsubsection{Existing Machine Learning Attack Methods}
Among existing modeling attack methods, some are tailor designed for the circuit architectures of the PUFs, including the popular attack on XOR PUFs (XPUFs) of R\"{u}hrmair et al. \cite{ruhrmair2010modeling} which employs the logistic regression (LR) for each component APUF of the XOR PUF, which has also been used in combination with other techniques for attacking the noise bifurcation protocol \cite{tobisch2015scaling} and the IPUF \cite{Wisiol2020Splitting}.
Some other attack methods employ generally applicable machine learning methods, like neural network methods \cite{Hospodar2012machine,aseeri2018machine,Santikellur2019deep,mursi2020fast,Thapaliya2021Machine,Wisiol2022IEEE}, which have cracked an extensive range of sophisticated PUFs, including XPUFs, FFPUFs, MPUFs \cite{sahoo2017multiplexer},  LSPUFs, IPUFs, FF-XPUFs \cite{Avvaru2020Homogeneous}.
%
%\subsubsection{The Machine Learning Attack Method Used in Our Experiments}
Among the generally applicable machine learning attack methods, the neural network method \cite{mursi2020fast} that employs hyperbolic tangent activation functions ($tanh$) for hidden layers exhibits the highest attack power for the most extensive range of PUFs, taking magnitudes lower training times and magnitudes fewer training data than earlier attack methods, even showed higher attack power on XPUFs than the LR-based method \cite{ruhrmair2010modeling} tailor-designed for attacking XPUFs. The attack power of this method \cite{mursi2020fast} on a range of PUFs was  confirmed by the study \cite{Wisiol2022IEEE}. Thus, the attack method \cite{mursi2020fast} is chosen for our experimental attack study of the interfaced PUFs.
%Thus, it is probably safe to say that, as of the time the paper is under preparation, the attack method \cite{mursi2020fast} is the most powerful broadly applicable machine learning attack method for delay-based strong PUFs.
\vspace{0.5mm}

\subsubsection{Parameters of Our Attack Method} $~$
%We choose the neural network-based attack method \cite{mursi2020fast} in our experimental attack study. %with parameters of the neural network chosen as follows.  
%
%To examine the effectiveness of the proposed interfaces, we carried out experimental attack studies on APUFs and XPUFs, and their interfaced  counterparts.   \cite{ruhrmair2010modeling,aseeri2018machine,mursi2020fast,Wisiol2022IEEE} show that neural network-based attack methods have the highest modeling power for capturing the behavior of a broad range of PUFs, we have decided to use the  for our attack study.

%The neural network attack method \cite{mursi2020fast} has the network structure as in Fig.\,\ref{fig:FF_NN.png}, with one input layer, followed by a feature generation layer, multiple hidden layers, and a single-node output layer.

\begin{figure}[!t]
 	\centering
 	\includegraphics[width = 3.3in]{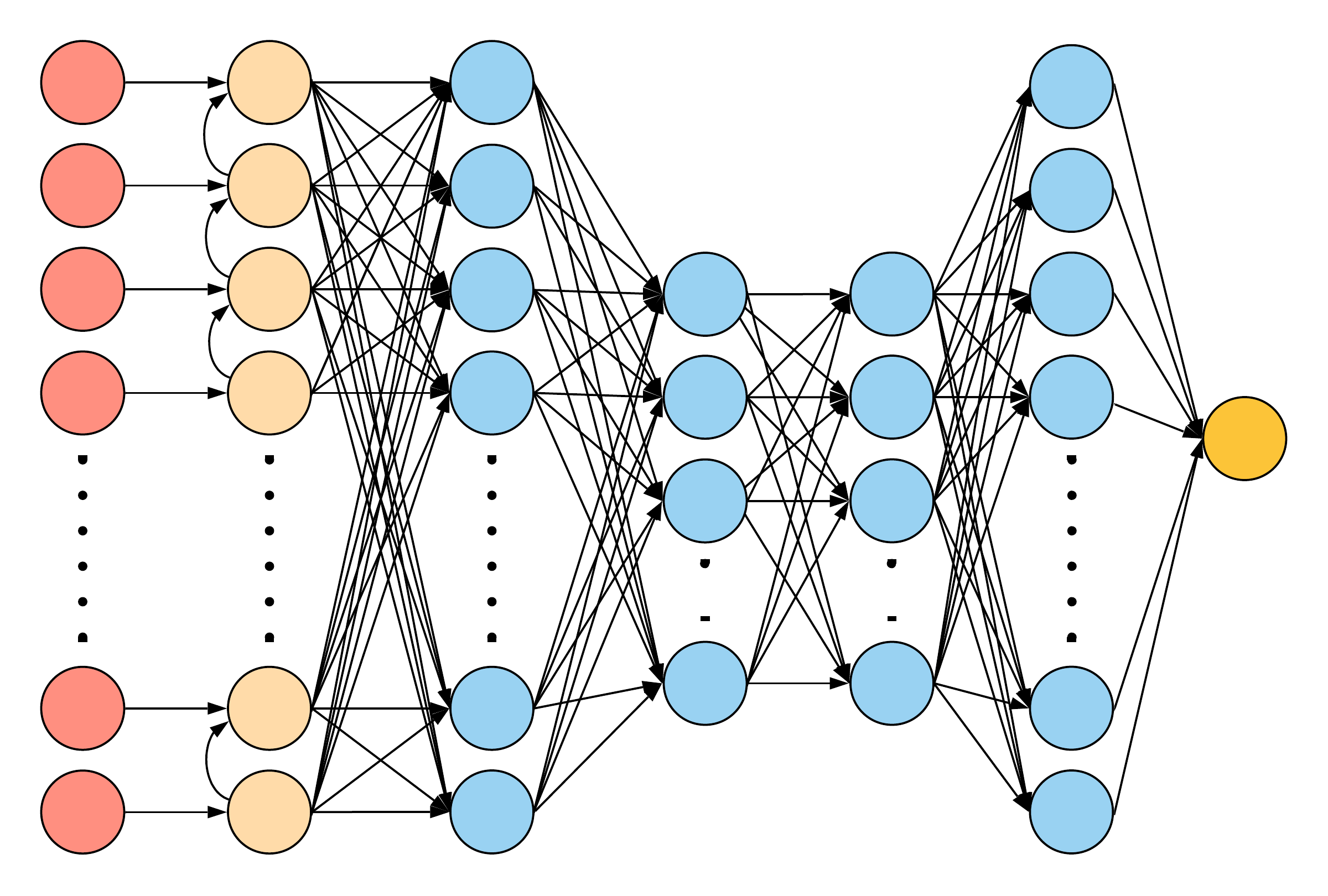}
 	\caption{The neural network architecture of the attack method}
 	\label{fig:FF_NN.png}
 \end{figure}

To choose values for parameters of the neural network attack method \cite{mursi2020fast} for our experimental attack study,
we ran test attacks using ranges of neural network parameter values on different PUF types, including the XOR PUF, the Interpose PUF, and the interfaced APUF, and choose   
%we iterated in the parameter space through multiple nested loops of parameter-choosing experiments,  with one loop iterating  through two activation functions (ReLU and hyperbolic tangent), the second loop iterating through numbers of hidden layers (from 3 to 6), and the remaining loops iterating through numbers of neurons on different layers. We ran these iterations on the APUF and interfaced APUFs with small $m$'s, and then a set of parameters that has the best balanced performances for all tested PUFs were chosen. The APUFs chosen in the experimental attack study  all have 64 stages. The chosen parameter values determine 
the following neural network architecture, with other parameters listed in Table~\ref{table_nn4ffpuf}.
%The chosen parameters determine the following network structure (also see Fig.\,\,\ref{fig:FF_NN.png} for illustration), with other parameters listed in Table~\ref{table_nn4ffpuf}.
 
\begin{itemize}
\item The input layer of $n+m$  bits $b_1, b_2,\cdots, b_{n+m}$, 
      where $n$ is the number of PUF stages and $m$ is the number of ghost bits;
\item the transform layer that transforms input bits from input layer  
      to $\phi$'s according to 
     
      $~~~~~~\phi(i) = (2b_i-1)(2b_{i+1}-1)\cdots(2b_{n+m}-1)$;
      
\item four layers of densely connected neurons whose 
      weights for all neurons are to be trained, where
      the numbers of neurons at these layers are specified in Table~\ref{table_nn4ffpuf}; %and remain the same for all PUFs, interfaced or not; and
\item the single-bit output layer to produce the output that models the response of the PUF.\vspace{0.5mm}
\end{itemize}

% \begin{figure}[]
% 	\centering
% 	\includegraphics[width = 3.35in]{FF_NN.png}
% 	\caption{Neural network attacking PUFs and their interfaced counterparts.}
% 	\label{fig:FF_NN.png}
% \end{figure}

\def\arraystretch{1.35}%  1 is the default
\begin{table}[ht] 
\caption{Neural Network for attacking 64-stage PUFs}
\centering
\begin{tabular}{|l|c|}
\hline
\multicolumn{1}
   {|c|}{{\bf{Parameter}}}  & {\bf{Description}} \\    \hline
   \textbf{Optimizing Method}     & ADAM                       \\    \hline
   \textbf{Hidden Lyr. Actv. Fx.} & tanh                       \\    \hline
   \textbf{Output Lyr. Actv. Fx.} & Sigmoid                    \\    \hline
   \textbf{Learning Rate}         & Adaptive                   \\    \hline
   \textbf{Hidden Layers}         & 4 hidden lyr. ($64, 32, 32, 64$) \\    \hline
   \textbf{Loss Function}         & Binary cross entropy       \\    \hline
   \textbf{Mini-batch Size}       & 100K                       \\    \hline
   \textbf{Kernel Initializer}    &Random Normal               \\    \hline
   \textbf{Epoch}                 & 500 \\    \hline
\textbf{Early Stopping} &  Validation accuracy $\ge$ 98\% \\    \hline
\end{tabular}
\label{table_nn4ffpuf}
\end{table}

\subsubsection{The Attack Software, Platform, and Set-ups}$~$

The attack method was implemented in Python using the TensorFlow machine learning library, and run on a single core of a multicore Intel Xeon Broadwell E5-2695 v4 processor of 2.1 GHz clock rate.
%and the available memory on the single node with 36 cores is 192 GB. 
The machine learning method for each PUF-interface instance is run for up to 500 epochs with an early stopping when the training validation accuracy 
reaches 98\%. The experiments used an 84-1-15 training-testing split, with 84\% of CRP data for training, 1\% of data for validation, and 15\% of the data for testing the trained model by predicting the response of the PUF using the trained model. 
%The codes we developed for the experimental study, including codes implementing the interfaces, 
%will be made available in the final version for reproducibility study by peer researchers.

%%%%%%%%%%%%%%%%%%%%%%%%%%%%%%%%%%%%%%%%%%%%%%%%%%%%%%%%%%%%%%%%%%%%%%
%\subsection{Comparative Attack Study on PUFs with or without the Interface}
\subsection{Comparative Experimental Attack Studies}

%%%%%%%%%%%%%%%%%%%%%%%%%%%%%%%%%%%%%%%%%%%%%%%%%%%%%%%%%

We have chosen a set of interfaces with different interface parameters for 64-stage APUFs. Interface notations with different interface parameter values as indicated  by the first column in Table~\ref{table_PlusResult}, where $G_m$ indicates interfaces with 18 ghost bits.
%where G-bits in the first column the table refers to ghost bits; and `` Separated'' in the column titled ``G-bits Property'' means that for all points $i_j$ in $G_m$  satisfies the condition $i_j+1 < i_{j+1}$ for $j=1,2,\cdots, m-1$ as given in Theorem~\ref{thm1}, and ``Contiguous'' means that $i_j+1 = i_{j+1}$ for all $j=1,2,\cdots, m-1$, which violates a condition in Theorem~\ref{thm1}. 
%The interface admits a permutation of the non-ghost bits before feeding them to PUF stages, and we did not study permuted non-ghost bits in Theorem~1 due to the difficulty to perform rigorous analysis of the effect of permuted non-ghost bits with the presence of ghost bits, so we did not examine them in our experiments.

In the experiments, for each type of interfaces listed in Table~\ref{table_PlusResult},  20 different instances of interface-PUF pairs were generated. In the generation of instances of interface-PUF pairs, for each interface-PUF pair, 
\begin{itemize}
\item interface %specification 
    parameters were chosen to satisfy the specifications of the interface type of the row, with some parameters randomly chosen (e.g. positions of ghost bits are random while meeting the requirements of the interface) and some other parameters selected uniquely (e.g. number of ghost bits), and
\item gate delay differences for APUF instances were randomly chosen.
\end{itemize}

\def\arraystretch{1.3}%  1 is the default
\begin{table}[ht]
\centering
\caption{Experimental Attack Results}
\label{table_PlusResult}
\begin{tabular}{|c|c|c|} \hline
%\begin{tabular}{|c|c|c|c|c|} \hline
%G-bits Property&{ $G_m$ }&{ CRPs}&{ Avg Acc}&{ Success Rate}\\ \hline
     PUF Type               &  Total CRPs &Success Rate\\  \hline
                            &     1 M     &  ~0\%  \\  \cline{2-3}
     9-XOR PUF              &     10 M    &  50\%  \\  \cline{2-3}
                            &     40 M    &  90\%  \\  \hline
                            &     1 M     &  ~0\%  \\  \cline{2-3}
(1,8)-Interpose PUF         &     10 M    &  10\%  \\  \cline{2-3}
                            &     40 M    &  90\%  \\  \hline
                            &     1 M     &  ~0\%  \\  \cline{2-3}
(7,7)-Interpose PUF         &     10 M    &  ~0\%  \\  \cline{2-3}
                            &     40 M    &  70\%  \\  \hline
%                            &     1 M     &  0\%  \\  \cline{2-3}
%$\rm G_{14}$-interfaced APUF&     10 M    &  ~5\%  \\  \cline{2-3}
%                            &     40 M    &  20\%  \\  \hline
                            &     1 M     &  ~0\%  \\  \cline{2-3}
$\rm G_{15}$-interfaced APUF&     10 M    &  ~5\%  \\  \cline{2-3}
                            &     40 M    &  10\%  \\  \hline
%                            &     1 M     &  ~0\%  \\  \cline{2-3}
%$\rm G_{17}$-interfaced APUF&     10 M    &  ~0\%  \\  \cline{2-3}
%                            &     40 M    &  10\%  \\  \hline
                            &     1 M     &  ~0\%  \\  \cline{2-3}
$\rm G_{18}$-interfaced APUF&     10 M    &  ~0\%  \\  \cline{2-3}
                            &     40 M    &  10\%  \\  \hline
                            &     1 M     &  ~0\%  \\  \cline{2-3}
$\rm G_{21}$-interfaced APUF&     10 M    &  ~0\%  \\  \cline{2-3}
                            &     40 M    &  ~0\%  \\  \hline
                            &     1 M     &  ~0\%  \\  \cline{2-3}
$\rm G_{24}$-interfaced APUF&     10 M    &  ~0\%  \\  \cline{2-3}
                            &     40 M    &  ~0\%  \\  \hline 
                            &     1 M     &  ~0\%  \\  \cline{2-3}
$\rm G_{27}$-interfaced APUF&     10 M    &  ~0\%  \\  \cline{2-3}
                            &     40 M    &  ~0\%  \\  \hline
\end{tabular}
\end{table}

In the experiments, from each instance of interface-PUF pair,  random CRPs were generated using the simulation software \cite{pypuf}. In each attack of a PUF-interface instance, the number of CRPs listed in a row in Table~\ref{table_PlusResult} is
the total number of CRPs used for the training, validation and 
testing, the Success Rate column lists the percentage
of attacks that produced testing accuracy of at least 80\%.
%and the Avg.\,\,Acc. column list the average testing accuracy of all attacks for the row, including both successful and unsuccessful attacks. 
An accuracy of 80\% is chosen as the threshold for the success of an attack, because PUF-based protocols employing APUFs or interfaced APUFs can easily use 90\% bit-matching as authentication threshold due to APUFs' high reliability usually at upper 90\%, much higher than the reliability of many sophisticated more modeling-attack-resistant PUFs (e.g. XPUFs, IPUFs). 
For such protocols, attacks with prediction accuracy below 80\% will have no chance to successfully impersonate a PUF. Besides, 80\% is a reasonably high standard for examining the security of an interface against modeling attacks since choosing 90\% will make a lot more interfaces look secure.

We carried out attacks on the computer described in Section~\ref{sec4_Experiment} and have listed The attack results in Table~\ref{table_PlusResult}. Due the high modeling-attack resistance of large XOR PUFs and IPUFs, for comparison, we also used neural networks of the same specifications as given in Table~\ref{table_nn4ffpuf} to attack some  XOR PUFs and IPUFs, and the attack results are also listed in Table~\ref{table_PlusResult}. 

The data in the table clearly show the effectiveness of the interface, and the  success rate of attacks on all tested interfaced APUFs were much lower than that of attacks on XPUFs and IPUFs.
In addition, the success rate for attacks on interfaced PUFs decreases with the increase of ghost bits, supporting Theorem~\ref{thm1} which reveals that the polynomial order of the separation surface of interfaced APUFs increases with the ghost bits.  
Attack success rate reaches 0\% success rate for all tested cases when the number of ghost bits hits 21 or higher, and these experimental attack results, together with Theorem~\ref{thm1}, indicate  that interfaced APUFs with ghost bits beyond 21 will very likely stay secure against modeling attacks. 

%The interface is the most effective for XOR\,PUFs.  The discussion after Theorem~1 shows that the order of the representing polynomial of an interfaced $x$-XOR PUF is $x$ times that of an interfaced APUF, and the experimental data support the analysis. 
%The data also showed that interfaces with consecutive ghost bits are weak against attacks (see Rows $B_{17,0}$, $B_{18,0}$, and $B_{20,0}$), which is consistent with Theorem~1 since it requires any two ghost bits be separated by a distance of 1 or larger. 
%
%Somewhat surprising to us is that the security of interfaced FF\,PUFs are weaker than that of 
%interfaced APUFs as indicated by the success rate, though the Avg.~ Acc. for tested 
%interfaced FF\,PUFs is not worse, and even slightly lower in many cases, than that of 
%interfaced APUFs. While we are not able to analyze the order of the representing polynomial of 
%%an interfaced  FF\,PUF, we expected them to be more secure than interfaced APUFs, because 
%un-interfaced FF\,PUFs have more complex challenge-response relationship than un-interfaced 
%APUFs have. But the experiments did not support our expectation.
%
%For interfaced APUFs, % and XPUFs, with separated ghost bits, 

The number of ghost bits needed to attain 0\% success rate is less than $n/2$ with $n$ being the number of PUF stages. Then, the data in Table~\ref{table_PlusResult} together with the discussion on issue (ii) in Sec.\,\ref{sec2.C} suggest that there are exponentially many interfaces which can secure 64-stage APUFs against the currently most powerful machine learning attack method.

%%%%%%%%%%%%%%%%%%%%%% End of Section 6 %%%%%%%%%%%%%%%%%%%%%%

\section{Conclusion}

Many IoT devices are resource-constrained, and call for authentication protocols implementable with low hardware overhead and operable with low power. Strong PUFs have the potential as hardware primitives for implementing lightweight authentication protocols, but many of them are vulnerable to modeling attacks. Modeling-attack-resistant PUFs are clearly more advantageous in enabling the development of simplistic authentication protocols with simple and lightweight operations and lower hardware overhead (when not counting the PUF), but most existing PUFs attain high security with large PUF circuit  architectural sizes, and large architectural sizes lead to higher hardware overhead. Ingeniously designed protocols exist that obfuscate the challenges, the responses, or both while using lightweight PUFs. 

In this paper, we are investigating the possibility of protocol-PUF co-design for a mutual authentication protocol using only probably the most lightweight strong PUF while maintaining hardware overhead at a level competitive with existing protocols of low overhead in both hardware and operation. The security of our protocol is designed to resist both conventional and reliability-based machine learning attacks, where reliability-based attacks are taken care of by protocol-level techniques and the resistance to conventional machine learning attacks is enabled by a zero-transistor challenge input interface which has a weakness protected by the protocol. 

The interface admits more input bits than the number of PUF stages, and these additional input bits are called ghost challenge bits.  When enough ghost bits are used, the challenge input interface has thwarted currently the most powerful modeling attack method in experimental attack studies, %by obfuscating the true challenge bits, 
and the experimentally observed growth in security against modeling attacks matches theoretical analysis which reveals that the polynomial order of the separation surface of the interfaced APUF increases with the ghost bits.
%when a condition on the ghost bits is met. 

With a low hardware-and-operation overhead and high security against modeling attacks, the co-designed protocol-PUF pair presents itself an excellent candidate for securely authenticating resource-constrained IoT devices.

\section*{Acknowledgement}
Computing resources at the High Performance Computing Center (HPCC) of Texas Tech University were used for part of the work.

\bibliographystyle{IEEEtran}

\bibliography{references}

 \begin{IEEEbiography}[{\includegraphics[width=1in,height=1.25in,clip,keepaspectratio]{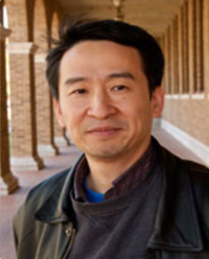}}]%
{Yu Zhuang}
received his PhD in Computer Science and PhD in Mathematics both in 2000 at Louisiana State University.  He was a visiting assistant professor at the computer science department of Illinois Institute of Technology from April to July of 2001, and has been with Texas Tech computer science department since September 2001. Dr. Zhuang's research interests include IoT security, high-dimensional data modeling and mining, high performance scientific computing. 
\end{IEEEbiography}

 \begin{IEEEbiography}[{\includegraphics[width=1in,height=1.25in,clip,keepaspectratio]{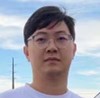}}]%
{Gaoxiang Li}
 received the B.S. degree in Computer Science and Technology from Shandong Normal University, Jinan, China, in 2017 and the M.S. degree in Computer Science from Auburn University, Auburn, USA, in 2020.  He is currently working toward the Ph.D. degree in Computer Science with the Department of Computer Science, Texas Tech University, Lubbock, USA. His research interests include machine learning, IoT security, and physical unclonable functions.
\end{IEEEbiography}

\end{document}